\documentclass{iau} 
\usepackage{graphicx}

\title[Remarkable AGN jets] 
{The remarkable AGN jets}

\author[S.Komissarov]   
{Serguei Komissarov
}

\affiliation{School of Mathematics, University of Leeds, \\ Leeds LS29JT,
United Kingdom \\ email: {\tt s.s.komissarov@leeds.ac.uk} }

\pubyear{2017}
\volume{324}  
\jname{New Frontiers in Black Hole Astrophysics}
\editors{A.C. Editor, B.D. Editor \& C.E. Editor, eds.}
\begin{document}

\maketitle

\begin{abstract}
The jets from active galactic nuclei exhibit stability which seems to be far superior compared 
to that of terrestrial and laboratory jets. They manage to propagate over distances up to a billion of initial jet radii. 
Yet this may not be an indication of some exotic physics but mainly a reflection of the specific environment these 
jets propagate through. The key property of this environment is a rapid decline of density and pressure along 
the jet, which promotes its rapid expansion. Such an expansion can suppress global instabilities, which require 
communication across the jet, and hence ensure its survival over huge distances.  At kpc scales, some AGN 
jets do show signs of strong instabilities and even turn into plumes. This could be a result of the flattening of the 
external pressure distribution in their host galaxies or inside the radio lobes. In this regard, we discuss the possible 
connection between the stability issue and the Fanaroff-Riley classification of extragalactic radio sources. 
The observations 
of AGN jets on sub-kpc scale do not seem to support their supposed lack of causal connectivity. When interpreted 
using simple kinematic models, they reveal a rather perplexing picture with more questions than 
answers on the jets dynamics.   

\end{abstract}

\firstsection 
\section{Introduction}

The terrestrial and laboratory gas and plasma jets are know to be susceptible to instabilities, which 
tend to destroy them over relatively short distances. This is particularly true for subsonic jets but also 
applies to supersonic jets. Although, supersonic jets are more resilient, the typical laboratory jets slow down and 
dissipate most of its kinetic energy over the distances which do not exceed a hundred initial jet radii. 
In contrast, the AGN jets appear to be much 
more robust. Indeed, the initial radius of an AGN jet is expected to be comparable to the gravitational radius 
of the central black hole, which is $r_i\sim 10^{14}\,$cm for $10^9 M_\odot$, and these jets can often 
be traced up to the distance of $100\,$kpc, which
is about one billion of the initial jet radius.  At first glance, this remarkable contrast between the AGN jets and their 
terrestrial counterparts suggests a dramatically different physics. After all, these jets are 
relativistic and produced by black holes.  We cannot reproduce them in a physical laboratory. However, the results of 
numerous theoretical and computational studies suggest that this may not be the right conclusion.

The properties of relativistic fluid jets are found to be quite similar to those of non-relativistic jets, when using proper
parametrisation. For example, initially it was claimed that high-Lorentz-factor jets are less unstable. However, 
both the relativistic Mach number and the specific inertia (mass density in the lab frame) depend on the Lorentz factor 
(\cite{K80, KF96})
and hence the results of the instability studies are well in line with the known dependence of growth rates on the 
Mach number and density ratio for non-relativistic jets. Moreover, the jets from young stars, which are non-relativistic, 
show a similar ability to propagate over huge distances, not much different from AGN jets when measured in 
the units of initial jet radius. 
The magnetic field can suppress some non-magnetic instabilities but introduces new current-driven instabilities, 
which could be even more destructive given its predominantly toroidal (azimuthal) structure expected far away 
from the magnetic central engine.       

The laboratory jets are approximately cylindrical (albeit with oscillations due to initial difference in pressure 
with the external gas) and most theoretical and numerical stability studies have been focused on cylindrical jets. 
This symmetry assumption allows
to simplify the problem and helps to make it treatable by standard analytic techniques.  
The AGN jets are not cylindrical, with the half-opening angle in the region of few degrees. This may not seem a lot and 
at first suggests that the cylindrical approximation may be suitable. On the other hand, over the long length of the jets, 
this amounts to a huge increase in  jet radius.  The same applies to the stellar jets. 
Although the stability of expanding jets is not well studied, it has been argued that such an expansion has a stabilising 
effect (e.g. \cite{MSO08}).     
 
 \section{Causality and jet stability}
 
It helps to distinguish between local and global instabilities of jets. Global modes imply coherent motion on the 
scale of the jet radius.  One typical example of such a global instability is the kink mode, where the whole jet moves 
sideways attaining a snake-like or a helical shape. Another is the pinch mode, where the jet radius varies independently 
of the azimuthal angle.  Such instabilities are recognised as the most threatening to the jet survival, as they lead to strong interaction 
with the surrounding gas in the non-linear regime.   Local instabilities are small scale motions. They may lead to local dissipation 
and mixing but do not endanger the jet integrity.  Clearly, only the global instabilities are relevant to the issue of the
astrophysical jets ability to propagate over very large distances.  

The very nature of global instabilities implies a causal communication across the jet. Without such a communication it is
impossible to support such highly-organised motions. In fluids, the communication speed is limited by 
the fundamental wave speeds, the  sound speed in gas dynamics and the fast magnetosonic speed in MHD. 
For supersonic (super-fast-magnetosonic) flows, these waves cannot propagate upstream. In fact, they are confined to a 
downstream-pointing cone known as the Mach cone. A higher Mach number corresponds to a narrower 
Mach cone and hence a longer distance over which the jet becomes aware of an upstream perturbation. 
This explains why the growth rate of instabilities in cylindrical jets decreases with the Mach number.         
In expanding supersonic jets, the Mach number monotonically increases and hence the growth rate decreases 
along the jet.  The instabilities become suppressed altogether when the typical opening angle of the Mach cone 
becomes smaller than that of the jet (This argument applies equally well to both relativistic and non-relativistic flows.).      

Consider for example a freely-expanding non-relativistic unmagnetised supersonic jet. Such a jet is almost conical and 
propagates at constant speed. Its density $\rho\propto z^{-2}$, pressure $p\propto \rho^\gamma$, sound speed 
$a^2\propto p/\rho$ and the Mach angle $\theta_m\propto a$. From these we find $\theta_m\propto z^{1-\gamma}$
and hence the Mach angle is bound to eventually become less than the opening angle of the jet, leading to its global 
stability.  Presumably, this simple argument was behind the assumption made in some early studies of kpc AGN jets that 
they emerge from AGN as cold conical flows (e.g. \cite[Falle 1991]{F91}).  
 
If a jet is not freely expanding but instead it is confined by an atmosphere with the power law pressure $p_{ext} \propto z^{-\kappa}$
then the ratio of the Mach angle and the jet opening angle $\theta_m/\theta_{j} \propto z^{(2-\kappa)/2}$ (\cite[Porth \& Komissarov 2015]{PK15}). 
This indicates that $\kappa=2$ is a critical value, with jets becoming causally-disconnected for $\kappa>2$. Such causally-disconnected 
jets are not necessarily freely expanding conical flows.  Indeed, in a conical jet its pressure decreases as $p\propto z^{-2\gamma}$, 
which is faster than that of the external gas if $\kappa<2\gamma$.  Hence for $2<\kappa<2\gamma$ the jets must still be 
confined by the external pressure, which is now communicated to the jet interior by shock waves.  
However, these are high amplitude waves and not the small perturbations 
associated with instabilities,  which suggests that for $\kappa>2$ jets are still globally stable.  Relativistic and magnetic 
effects do not seem to alter this conclusion  (\cite[Porth \& Komissarov 2015]{PK15}).  

The central engines of AGN and 
stellar jets are dominated by compact objects, whose gravitational field has a strong impact on the distribution of 
surrounding gas. The overall effect is a rapid decrease of  both its density and pressure away from the centre. 
For a polytropic atmosphere of central mass, one has $\kappa=\gamma/(\gamma-1)$,
which is higher than 2 when $1<\gamma<2$. For a spherical adiabatic wind, $\kappa=2\gamma$, which is also 
steeper than the critical one.  A self-collimating magnetic wind can in principle deliver $\kappa<2$. 
For the Bondi accretion $\kappa =3\gamma/2$, which is still larger than 2 for 
$\gamma>4/3$. Thus, steep gradients of external pressure, bordering the critical 
value, are expected to be quite common close to the central engine. Unfortunately, we are 
still unable to measure the parameters of the external gas on the scales below $1\,$kpc.  Instead, we have to rely mostly 
on indirect model-dependent estimates.  Taken together, this circumstantial  evidence suggests that 
$\kappa\simeq2$ is indeed quite typical for the AGN 
environment (e.g. \cite[Phinney 1983, Begelman et al. 1984]{phinney-83,BBR-84}). 

\begin{figure}[h]
\begin{center}
 \includegraphics[height=2.in]{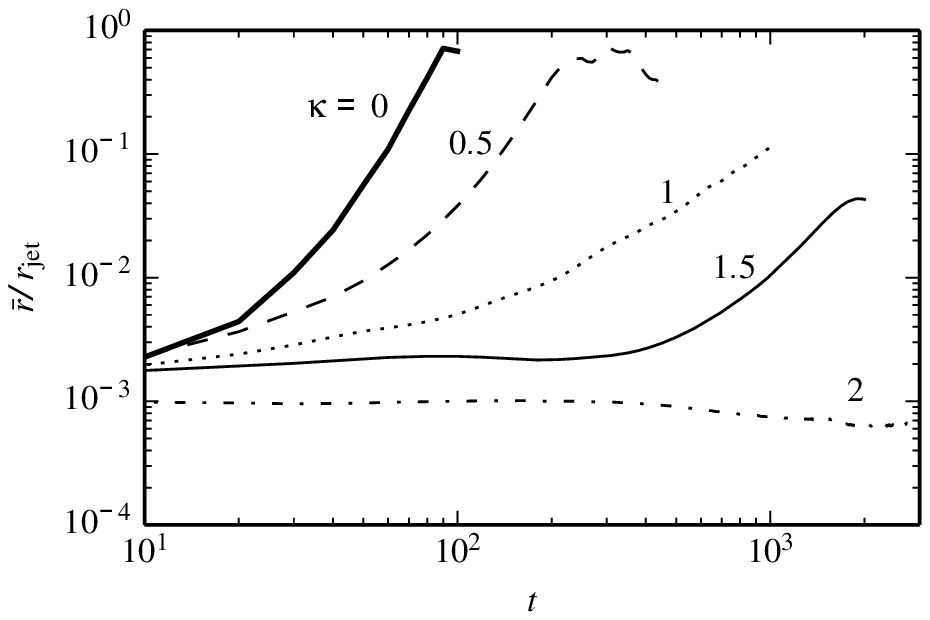} 
 \includegraphics[height=2.in]{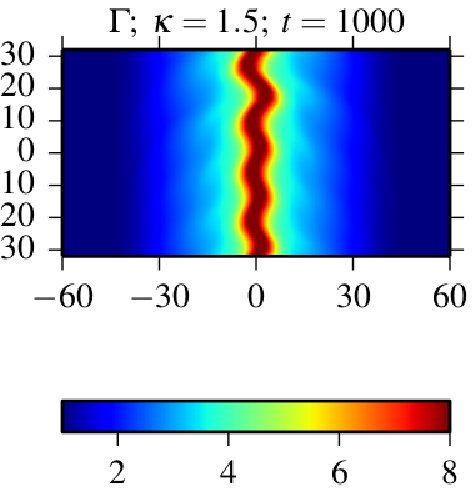} 
 \caption{\textit{Left panel}: Amplitude evolution of a kink-mode instability as observed in the simulations by \cite{PK15}. 
 The amplitude is defined as the sideways displacement of the jet barycentre measured in the units 
 of the instantaneous jet radius. The time is measured in the units of $r_0/c$, where $r_0$ is the initial 
 jet radius. \textit{Right panel}: The unstable core of the jet in the atmosphere with $\kappa=3/2$. The image shows 
 the Lorentz factor distribution. 
 }
   \label{sk-fig1}
\end{center}
\end{figure}
 
In order to study the effect of jet expansion on its stability, \cite{PK15} used periodic box simulations where the 
pressure of external gas was forced to decrease as a power function of time, similarly to what would have 
been seen in a frame moving radially at constant speed through a power-law atmosphere. 
The left panel of Fig.\,\ref{sk-fig1} shows the 
growth of a kink-mode perturbation in a moderately relativistic magnetised jet with a purely azimuthal magnetic 
field.  One can see, that the growth rate decreases with $\kappa$ and the instability is totally suppressed for 
$\kappa=2$, in line with the expectations.

One characteristic property of magnetised jets is the development of a magnetically-confined core which refuses to 
expand with the same rate as the rest of the jet and remains causally connected. Hence, this core can develop 
instabilities and gets destroyed even when the rest of the jet remains stable. In fact,  this is an example of a local 
instability at work. 
The right panel of Fig.\,\ref{sk-fig1} shows how this instability develops in the simulations by \cite{PK15}.
Such local instabilities may power emission from jets without endangering their integrity.   
  
\section{The Fanaroff-Riley division}

The issue of jet stability is often discussed in connection to the global division of extragalactic radio sources into 
two main classes according to the classification by \cite{FR74}. The main motivation 
behind this connection is based on the remarkable similarity in appearance between the kpc-scale jets (and 
tails) of some FR1-type sources (e.g. 3C31) and the unstable transonic jets produced in laboratory. 
These sources do not show any cocoons separating their jets from the external gas and in some cases this
gas is observed in X-rays  (e.g. \cite[Schreier et al. 1982]{SGF82} for M87). Furthermore, 
the theoretical model based on the idea of a turbulent flow confined by the external pressure and entraining 
significant  amount of surrounding gas has successfully reproduced a number of the observational results 
(\cite[e.g. Bicknell 1984, Komissarov 1990, Laing \& Bridle 2002]{B84,K90,LB02}).  
Thus, we need to take seriously the possibility that some AGN jets become reconfined by the external gas pressure 
at the distances of $0.1\div1\,$kpc from the central engine and develop instabilities.  
In fact, the optical observations of giant elliptical galaxies indicate that their stellar distribution flattens near the centre. 
The size of this central core depends on the galactic mass and is scattered around 1 kpc.  At such distances,  
the gravitational potential of the central black hole is sub-dominant and the gas distribution is expected to reflect 
that of the stars. In fact, it is known that outside of the core the gas distribution does mimic the stellar one    
 (\cite[Mathews \& Brighenti 2003]{MB03}).  Hence, the X-ray gas of galactic coronas is likely to have
 a central core with flat pressure distribution which could be sufficiently high to reconfine the FR1-type jets.  
 
The reconfinement process involves a strong shock, driven into the jet by the external pressure. 
It turns out (see \cite[Porth \& Komissarov 2015]{PK15}) that for a uniform external gas of 
pressure $p_{0}$ and an initially conical jet of kinetic power $L_{j}$, the reconfinement shock reaches the jet 
axis at the distance $z_{r,kpc} \approx 0.3 (L_{j,44}/p_{0,-9})^{1/2}$, where the units are $1\,$kpc, 
$10^{44}\,$erg/s and $10^{-9}\,$dyn/cm$^2$.  Since $p_{0,-9}\approx1$ is typical for the central parts of 
the X-ray coronas of elliptical galaxies and for FR1 sources $L_{j,44}<1$ (\cite[Cavagnolo et al. 2010]{CMN10}), 
the reconfinement of FR1 jets inside the galactic cores seems very likely. On the other hand, the kinetic power of FR2 
is significantly higher and  the reconfinement process cannot be completed inside the cores.  Outside,
the pressure of the X-ray gas decreases rapidly again, $p\propto z^{-1.5\pm0.25}$ 
(\cite[Mathews \& Brighenti 2003]{MB03}). This slows down not only the reconfinement process but also 
the development of instabilities that follows. So it is reasonable to conclude
that jets which manage to escape the core will remain as stable supersonic flows for much longer.      

In fact, the observations show that FR1 and FR2 sources belong to different regions in the $P-L_O$ plane, 
where $P$ is the source radio power and $L_O$ is the optical luminosity of the host galaxy, with the 
dividing line $P^{FR}\propto L_O^2$ (\cite[Owen \& Ledlow 1994]{OL94}). Assuming that the critical jet 
power, $L_j^{FR}$, separating FR1 and FR2 sources is set by the condition $z_r=z_c$, where $z_c$ is the 
X-ray core radius, \cite{PK15} obtain $L_{j}^{FR} \propto p_0 z_c^2$.  Thus, the Owen-Ledlow result can be 
explained if the parameters of X-ray cores depend on the optical luminosity (mass) of the parent galaxy 
in an agreeable way.  Unfortunately, the X-ray cores are rarely resolved. If, however, their radius scales in the same 
way as that of the stellar distribution then $z_c\propto L_O^{1.1}$ and if the gas temperature scales as the stellar 
velocity dispersion then $p_0\propto L_O^{-0.15}$ (\cite[Porth \& Komissarov 2015]{PK15}). 
This leads to $L_{j}^{FR} \propto L_O^{2.05}$, 
which is basically the same as observed, provided $P \propto L_j$. Overall, this is a very encouraging result.

Contrary to being often described as the prototype FR1 source, 
3C31 represents only a small sub-class of FR1 sources. Most of them have jets submerged into radio lobes, which  
makes them more like FR2 sources. The difference in that in FR2 sources the jets remain well collimated all the way
and terminate in a very bright and compact hot spot near the outer edge of the radio lobe, whereas the jets of FR1 sources 
seem to ``disappear`` half-way through the lobe (\cite[de Ruiter et al. 1990]{dR90}). These observations show that 
the model of a naked jet has its limitations and a better model of FR-1 sources has to incorporate not only the conditions 
close to the AGN, where their jets are naked, but also the distant regions where these jets inflate radio-lobes.

The widely accepted current model of FR2 radio sources involves two jets surrounded by a light cocoon, which 
shields them 
from direct contact with the interstellar/intergalactic gas (\cite[Falle 1991, Komissarov \& Falle 1998, 
Kaiser \& Alexander 1997]{F91,KF98,KA97}). This cocoon is filled with plasma which has been supplied
by the jets during their previous activity and passed through the jet termination shocks, associated in the model with 
the leading hot spots of a FR2 source. This cocoon is over-pressured  with respect to the external gas and 
drives a strong shock through it.  The expansion of this cocoon is slow compared to its sound speed and 
hence its pressure distribution is more or less uniform. 

\cite{F91} has shown that the cocoon pressure 
is always sufficient to reconfine the initially free conical AGN jet and turn it into a quasi-cylindrical flow whose length/radius 
ratio increases with time.  \cite{F91} argued that eventually the cylindrical section becomes long enough for the jets 
to develop instabilities and become turbulent, now inside the cocoon. He proposed that this may result in the change 
in morphology from the FR2 to the FR1 type. However his 2D axisymmetric simulations could not confirm this prediction.  
The 2D simulations of jets with slab symmetry by \cite{KF03} did exhibit a kink-mode instability but the lobe 
structure still remained more or less unchanged. The expected transition has been achieved only in the very 
recent 3D hydrodynamic simulations by \cite{MBR16} and RMHD simulations by \cite{TB16}.  It turns out that the 3D effects 
are critical in this problem.  While an interesting and promising idea, it remains to be seen if it can explain the 
separation between FR classes in the Owen-Ledlow plane. 

The analytical model of FR2 sources by \cite{F91} deals only with power-law density distributions of 
the external gas as this assumption leads to a self-similar solution.  In the solution the whole jet is surrounded 
by a melon-shaped expanding cocoon.  
This naturally leads to a picture where the interstellar gas is evacuated from the parent galaxy. However, this 
is likely to be an over-simplification. From the theoretical viewpoint, an external gas distribution with flat central core 
introduces a characteristic length-scale to the problem and this is likely to alter the dynamics.  
Observationally, the two-lobed structure of FR-2 radio-sources, even at low radio frequency where the synchrotron losses are week 
(e.g. \cite[Kassim et al. 1996]{KPC96}), suggests that the melon-shape of their cocoons is not generic.  
This conclusion is strengthened by the X-ray observations which show the presence of hot gas in the gap
between the lobes of FR2 sources (\cite[Nulsen et al. 2015]{NYK15}). 
Thus, it looks more likely that close to AGN the FR2 jets are ``naked'' and only further out they are cocooned.

\section{Complications}

The idea that the rapid expansion of AGN jets  delays the onset of global instabilities until 
kpc scales offers a simple and attractive explanation to the most basic observed properties of extended 
extragalactic radio sources within the framework of well known physics.  Yet we need to learn more about the 
properties of ISM and the dynamics of AGN jets on sub-kpc scales in order to see if these jets 
are indeed largely causally-disconnected on these scales. 

Using the basic steady-state theoretical models, one may expect such flows to have a relatively simple structure. 
Not affected by global instabilities, they should be more or less straight.  In the absence of strong internal dissipation
of kinetic energy, their acceleration should be quite efficient  and so we expect them to be highly supersonic.  
Even for the rather slow magnetic collimation-acceleration mechanism, the acceleration should be completed by 
the distances corresponding to the blazar zone, $z \lesssim 1\,$pc  (\cite[Komissarov et al. 2007]{KBV07}). 
Their shape is not that well determined.  Although, highly-magnetised causally-disconnected jets are  
almost conical (\cite[Komissarov et al. 2009]{KVK09}), the AGN jet are not expected to remain highly-magnetised 
on the distances above one parsec. As to weakly-magnetised jets, we have already noted that they cannot be conical 
unless $\kappa > 2\gamma$\footnote{For very hot jets with $p\gg\rho c^2$ this condition reads $\kappa>4$ but 
jets which have reached their asymptotic speed cannot remain that hot. }.   Assuming that the dissipation associated 
with the shocks which adjust the jet pressure to that of the external gas with $<2\kappa<2\gamma$ is weak and they 
can still be treated as adiabatic, we find that the jet radius  $r_j\propto z^{\kappa/2\gamma}$.  So the shape is still parabolic.
Weak oscillations associated with stationary shocks may be superimposed on top of this parabolic shape throughout 
the jet length. 
We also expect to find $\theta_j/\theta_m >1$. For magetically-dominated flows $\theta_m \approx \sigma^{1/2}/\Gamma$, 
where $\sigma\gg1$ is the relativistic magnetisation parameter, whereas for particle-dominated plasma ($\sigma\ll1$)
$\theta_m \approx \beta_s/\Gamma$, where $\beta_s=a/c$ is the sound speed. Hence the condition of 
causal disconnection reads as $\theta_j\Gamma>\sigma^{1/2}$ and $\theta_j\Gamma>\beta_s$ for these two
regimes respectively.  

The properties of AGN jets on sub-kpc can be studied using the VLBI method; normally down to about one parsec from 
the central engine. These observations allow to measure the jet apparent opening angle and the apparent velocity of 
individual components in the plane of the sky. To convert these into the actual values one has to know the angle 
between the jet and the line of sight. A number of more or less reasonable assumptions can be made to estimate 
the viewing angle.  For individual sources they may lead to significant errors but for a large sample one expects to 
obtain statistically reliable results.  Analysis of large samples has become the main approach of the current studies 
of VLBI-jets.  Several groups studied the distributions of $\Gamma$ and $\theta_j$ using somewhat different 
models and techniques  and they all agreed that $\theta_j \Gamma= 0.1\div0.2$ 
(\cite{JML05,PKL09,CSP13}).  This means that the jets can be causally disconnected only if they are 
particle-dominated and cool, with the sound speed $\beta_s \le 0.1$.  This implies that no matter what is 
the main energy type of the jets near the central source by the distance of about one parsec it is converted 
into the kinetic energy of bulk motion. This is consistent with the conclusion reached by \cite{KBV07} that the 
magnetic acceleration of AGN jets should be completed inside the  blazar zone.  The alternative 
mechanism of thermal acceleration is even faster, with $\Gamma\propto r_j$.    

However, \cite{LAA13} have found that in the MOJAVE sample the jet components identified at larger 
projected distance $z_{p}$ from the core tend to have higher apparent speed.  Moreover, 
\cite[Homan et al.(2015)]{HLK15} have 
shown that individual components move with predominantly positive acceleration at $z_{p}< 4\,$pc and 
predominantly negative accelerations at $z_{p}>30\,$pc. 
If these proper motions are indicative of the bulk flow velocity then 
in stark conflict with the current theoretical models the jet speed keeps increasing
even on the scales well above $z=1\,$pc and hence even at such large distances
there is still plenty of free energy in the magnetic or  thermal form to power this acceleration. 
            
The M87 jet stands out from the rest by the fact that this galaxy is very close and the jet can be traced 
almost all the way from the event horizon to the kpc scales.  According to the observations, 
there exist three distinct zones. In the \textit{near zone}, which extends up to about $z=0.1\,$pc ($100\,r_s$), 
the jet shape is approximately parabolic with $r_j\propto z^{0.75}$ 
(\cite[Hada et al. 2013]{HKD13}). Here the jet is still inside the AGN.   The \textit{middle zone} 
extends up  $z=300\,$pc. In this zone, the jet shape is a different parabola with $r_j\propto z^{0.56}$ 
(\cite[Hada et al. 2013]{HKD13}). 
This shape change may indicate a change in the jet environment. Presumably, in the middle  
zone the jet is already surrounded by ISM of the parent galaxy\footnote{Using the theoretical shape 
$r_j\propto z^{\kappa/2\gamma}$ and assuming $\gamma=5/3$, we find $\kappa=2.5$ and $1.87$ for 
the near and middle zones respectively. These values seem quite reasonable.}.  The \textit{far zone} extends into kiloparsecs
and the jet shape is almost conical there (e.g. \cite[Nakamura \&Asada 2014]{NA14}). 

The apparent speed of individual features is seen to increase with distance in the middle zone 
from the Lorentz factor $\Gamma\approx 1$ at $z=50\,$pc to $\Gamma\approx 6$ at $z=300\,$pc 
(\cite[Asada et al. 2014]{AND14}).  \cite{KLH07} report even smaller speeds, below few percent 
of the speed of light, for $1\,$pc$<z<50\,$pc. Optical observations indicate that in the far zone the apparent 
speed of proper motion decreases with distance. However,  the velocity pattern is extremely complicated, 
with very large local variations within individual knots (\cite{MSB13}). If taken as measurements of the bulk 
speed of a quasi-steady jet, the results by \cite{AND14} and \cite{KLH07} indicate that up to the distance 
$z\approx 50\,$pc the jet is still subsonic! Indeed, in the magnetic model the fact that the jet 
eventually reaches $\Gamma\approx 6$
means that at the start of the acceleration zone the flow magnetisation parameter $\sigma>6$. This 
corresponds to the Alfv\'en speed $c_a>0.91c$. In the thermal model, this 
means that at the start of the zone the jet plasma is ultra-relativistically hot with the sound 
speed  $a\approx c/\sqrt{3}$. Both the numbers are high enough to imply a subsonic motion at $z<50\,$pc, 
which is also in stark conflict with our current understanding of AGN jets.  The low speeds also imply the Doppler 
factor of order unity, which suggests that the M87 jet is intrinsically one-sided and hence rather exceptional. 
Interestingly and in conflict with the above results, \cite{WLJ08} report the apparent speed $\approx 2c$ at $z\approx 1\,$pc.
Perhaps the interpretation of VLBI observations is not that straightforward and the observed proper motions
are poor indicators of the jet speed.

Idealised steady-state theoretical models of jets have a very simple structure, either entirely featureless or with few
clearly identifiable reconfinement shocks. Even a quick glance at the high-resolution images of AGN 
jets reveal that their structure is much richer.  Variable central engine, cloudy environment
and instabilities are the most obvious factors which can be responsible for this. All these factors can
strongly modify the jet dynamics and make the idealised models redundant.  
More adequate models can be based on advanced computer simulations which have to accommodate the 
3D nature of AGN jets. In these models, the issue of non-thermal particle acceleration and emission has to 
be addressed with rigour for the sake of direct comparison with the observations.  
Although some may think that we have captured the essence of the AGN jets, observations keep asking us 
challenging questions about this remarkable astrophysical phenomenon and force us to have another look at the 
basics again and again. We may well be just at the beginning of a long and complicated journey.




\end{document}